# Ultrafast optical properties of stoichiometric and non-stoichiometric refractory metal nitrides TiN$_x$, ZrN$_x$, and HfN$_x$

Jarosław Judek,[1,*] Rakesh Dhama,[2] Alessandro Pianelli,[2] Piotr Wróbel,[3] Paweł Piotr Michałowski,[4] Jayanta Dana,[2] and Humeyra Caglayan[2]

[1] Institute of Microelectronics and Optoelectronics, Warsaw University of Technology, Koszykowa 75, 00-662 Warsaw, Poland.

[2] Faculty of Engineering and Natural Sciences, Tampere University, Tampere 33720, Finland.

[3] Faculty of Physics, University of Warsaw, Pasteura 5, 02-093 Warsaw, Poland.

[4] Łukasiewicz Research Network – Institute of Microelectronics and Photonics, Al. Lotnikow 32/46, 02-668 Warsaw, Poland

* Corresponding author, e-mail address: jaroslaw.judek@pw.edu.pl.

**Abstract:** Refractory metal nitrides have recently gained attention in various fields of modern photonics due to their cheap and robust production technology, silicon-technology compatibility, high thermal and mechanical resistance, and competitive optical characteristics in comparison to typical plasmonic materials like gold and silver. In this work, we demonstrate that by varying the stoichiometry of sputtered nitride films, both static and ultrafast optical responses of refractory metal nitrides can efficiently be controlled. We further prove that the spectral changes in ultrafast transient response are directly related to the position of the epsilon-near-zero region. At the same time, the analysis of the temporal dynamics allows us to identify three time components - the "fast" femtosecond one, the "moderate" picosecond one, and the "slow" at the nanosecond time scale. We also find out that the non-stoichiometry does not significantly decrease the recovery time of the reflectance value. Our results show the strong electron-phonon coupling and reveal the importance of both the electron and lattice temperature-induced changes in the permittivity near the ENZ region and the thermal origin of the long tail in the transient optical response of refractory nitrides.

1. Introduction

Refractory metal nitrides, particularly group IVb metal nitrides like TiN, ZrN, and HfN, are well-known materials with an established position in the industry due to their robust mechanical and electrical properties. Nevertheless, current scientific interest focuses more on their optical properties [1] since they pose decent plasmonic properties and thus are considered a promising alternative to gold or silver [2]. The main advantages of refractory metals over noble metals lie in their compatibility with silicon technology, as well as their thermal, mechanical, and chemical endurance, and also in robust and cheap fabrication technology.

Group IVb metal nitrides, at first sight, are considered to be worse plasmonic materials in comparison to gold or silver since they are characterized by the less negative real part and the more positive imaginary part of the dielectric function. However, it turns out that despite theoretically inferior properties, metal nitrides can outperform gold and silver in many practical plasmonic applications. For example, titanium nitride nanodisks surpass their gold counterparts as efficient local heat sources in the biological transparency window [3]. Titanium nitride nanoparticles dispersed in water exhibit lossy plasmonic resonance enabling efficient absorption of sunlight and superior performance to heat water and generate vapor [4]. TiN nanoantennas examined for nonlinear plasmonic application outmatch gold counterparts by one order of magnitude regarding laser power sustainability [5]. Hot carrier photodetector based on TiN generates significantly higher photocurrent than that obtained using gold in a similar structure [6]. Plasmonic TiN nanoparticles on TiO2 nanowires enhanced solar water splitting with respect to Au counterpart [7]. The TiN, ZrN, and HfN nanocrystals demonstrated higher photothermal transduction efficiencies than an Au nanorod benchmark [8]. While the high broadband reflectivity mechanical, structural, and thermal stability of HfN and ZrN made them an attractive alternative for solar mirrors [9].

Recently, TiN has been combined with transparent conducting oxides (TCOs) to design and realize multilayered meta-devices; such meta-structures can enable two epsilon-near-zero regions (ENZ) in the visible range as well as in near-infrared range (NIR) range along with ultrafast switching responses [10]. Additionally, a single device based on a bilayer of nitride and TCO film is reported to investigate wavelength-dependent "fast" and "slow" time responses [11].

Different classes of plasmonic materials demonstrate different temporal dynamics upon femtosecond excitation. For example, plasmonic noble metals, particularly gold, exhibit temporal response only in the picosecond timescale and render the hot-electrons thermalization process due to the electron-electron and electron-phonon scattering [12–14]. Similar but slightly faster kinetics were observed in transparent conductive oxides like ITO and AZO [15, 16]. But in the case of TiN, ZrN, and HfN [17–22] consequently demonstrate that there are typically three distinguishable components in the kinetics of photoexcited charge carriers, occurring on different time scales – one in the femtosecond range, another in the picosecond range, and third in the nanosecond time scale. These components have recently been identified as electronic, thermoelastic, and thermal origin [18]. However, we note that there is also a different interpretation provided based on the results obtained for hafnium nitride thin films – that the long-lasting changes in transient optical properties are due to long-living hot carriers, which is suspected to be the results of restricted phonon decay pathways due to a large gap in phonon band structure [20–22]. Another recent report on ultrafast transient absorption on HfN nanoparticles suggests an ultrafast light conversion into heat with no sign of long-living [23].

The optical properties of refractory metal nitrides can be easily and efficiently controlled or tuned by stoichiometry. Notably, the ENZ wavelength in TiN [24–27], ZrN [28–31], and HfN [32, 33] depends on the metal-to-nitrogen ratio. Tuning the optical properties is beneficial when considering future photonic applications that require operation at a specific wavelength that does not precisely match the wavelength offered by stoichiometric materials. In this work, we investigate the titanium, zirconium, and hafnium nitrides and demonstrate the possibility of tuning the ultrafast properties of refractory nitrides in a spectral domain by stoichiometry control, which is an easy and reliable operation from a technological point of view (e.g., in the reactive magnetron sputtering technique, the stoichiometry of the deposited films depends directly on the nitrogen flow, while keeping all the rest parameters constant). Furthermore, we analyze how spectral and temporal properties of broadband ultrafast transient reflectance change with the films' composition.

## 2. Methods

All samples examined in this study, i.e., the titanium, zirconium, and hafnium nitrides, were deposited by a reactive magnetron sputtering technique using Oxford Instruments Plasmalab System 400 from pure titanium, zirconium, and hafnium targets at room temperature on a silicon substrate. Thus, all the samples are polycrystalline with a typical columnar microstructure. Applied power, pressure, and deposition time were constant for all the samples, and the stoichiometry was controlled by the nitrogen flow while keeping the Ar flow constant. It is a typical well-established approach for thin film deposition with a certain nitrogen-to-metal ratio.

Static optical properties, namely the real and imaginary parts of the dielectric function, were obtained from Variable Angle Spectroscopic Ellipsometry (Woollam RC2 ellipsometer) in the spectral range of 193 nm – 1690 nm. The stoichiometry of $TiN_x$, $ZrN_x$, and $HfN_x$ samples was determined using the SIMS technique (IMS SC Ultra instrument, CAMECA, Gennevilliers Cedex, France) under ultra-high vacuum, typically $4 \times 10^{-10}$ mbar.

To investigate the transient response of refractory metal nitride films, ultrafast time-resolved pump-probe spectroscopy was employed in reflection mode. The femtosecond pulsed Ti: Sapphire laser (Libra F, Coherent Inc.), with a wavelength of 800 nm, a pulse width of approximately 100 fs, and a repetition rate of 1 kHz, enables the fundamental laser pulses. In our case, 90 % of the fundamental laser beam was directed towards an optical parametric amplifier (OPA) (Topas C, Light Conversion Ltd.) to generate the desired excitation wavelength of 400 nm. Such resulting excitation beam with a diameter of 1 mm was attenuated to an energy density of 400 μJ cm$^{-2}$ to pump all the nitride samples. The remaining 10 % of the fundamental laser beam travels toward a motorized stage (delay line) to enable the delay between pump and probe pulses and then subsequently focused onto a 2 mm cuvette with water for generating continuum white light (400 nm - 800 nm) for the probe light. This probe beam was then split into reference and signal beams. The transient reflectance change was measured in chopper mode synchronized with fundamental laser pulses. The spectra averaged over 3000 excited pulses for each delay time value. An Exci-Pro spectrometer (CDP Inc.) equipped with a CCD array for the visible spectral range (400 nm – 750 nm) was used to record transient reflectance in delay up to 6 ns.

## 3. Results and Discussion

Table 1. Composition and selected properties characterizing stationary optical properties.

|   | titanium nitride | | | | zirconium nitride | | | | hafnium nitride | | | |
|---|---|---|---|---|---|---|---|---|---|---|---|---|
|   | $x$ in $TiN_x$ | $\lambda@\varepsilon_1=0$ (nm) | $\varepsilon_2$ | FoM | $x$ in $ZrN_x$ | $\lambda@\varepsilon_1=0$ (nm) | $\varepsilon_2$ | FoM | $x$ in $HfN_x$ | $\lambda@\varepsilon_1=0$ (nm) | $\varepsilon_2$ | FoM |
| 1 | 0.93 ± 0.02 | 428 | 4.1 | 1.13 | 0.93 ± 0.03 | 344 | 4.5 | 0.4 | 0.90 ± 0.03 | 360 | 2.5 | 1.30 |
| 2 | 0.99 ± 0.01 | 479 | 3.1 | 1.86 | 1.01 ± 0.01 | 409 | 2.3 | 2.1 | 0.98 ± 0.01 | 378 | 2.2 | 1.82 |
| 3 | 1.04 ± 0.01 | 501 | 4.3 | 1.59 | 1.06 ± 0.02 | 419 | 2.5 | 1.9 | 1.04 ± 0.01 | 396 | 2.3 | 1.71 |
| 4 | 1.10 ± 0.01 | 537 | 4.9 | 1.30 | 1.10 ± 0.02 | 439 | 2.9 | 1.6 | 1.10 ± 0.02 | 422 | 2.6 | 1.45 |
| 5 | 1.16 ± 0.02 | 572 | 5.2 | 1.14 | 1.14 ± 0.02 | 489 | 3.7 | 0.9 | 1.16 ± 0.02 | 448 | 2.9 | 1.17 |
| 6 | 1.21 ± 0.02 | 608 | 5.6 | 0.99 | 1.19 ± 0.03 | 542 | 4.5 | 0.6 | 1.21 ± 0.03 | 497 | 3.6 | 0.77 |

Table 1 shows selected properties characterizing the stationary optical properties of examined refractory nitride films – the correlation between composition expressed as the nitrogen-to-metal ratio and the wave-length at which the real part of the permittivity equals zero $\lambda@\varepsilon_1=0$, the value of the imaginary part of the permittivity at the wavelength mentioned above $\varepsilon_2$, and the maximum value of the plasmonic Figure of Merit (FoM) defined as a minus ratio of the real to the imaginary part of the permittivity FoM = $-\varepsilon_1/\varepsilon_2$. These results prove that changing the stoichiometry to approximately ± 20 % can change the $\lambda@\varepsilon_1=0$ by about 150 nm – 200 nm. The cost is the higher losses expressed as the $\varepsilon_2$ value and lower FoM. Our results are in line with other literature reports both qualitatively and quantitatively. Figure 1 shows curves of the real $\varepsilon_1$ and imaginary $\varepsilon_2$ part of the dielectric function, the plasmonic FoM = $-\varepsilon_1/\varepsilon_2$, and simulated reflectance spectra of stoichiometric and non-stoichiometric titanium, zirconium, and hafnium nitrides as a function of wavelength.

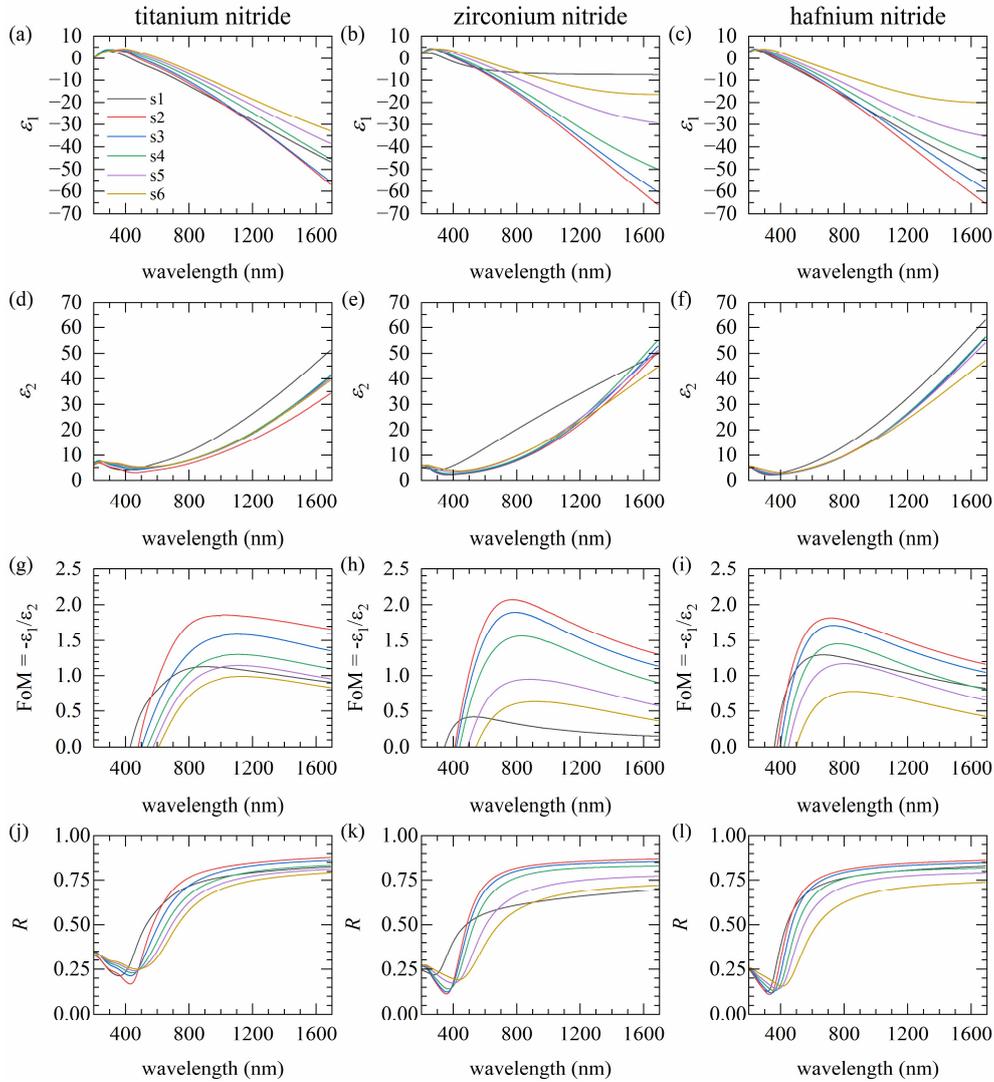

Fig. 1. Optical properties – the real $\varepsilon_1$ and imaginary $\varepsilon_2$ part of the dielectric function, the plasmonic Figure of Merti FoM = $-\varepsilon_1/\varepsilon_2$, and simulated reflectance spectra of stoichiometric and non-stoichiometric titanium, zirconium, and hafnium nitrides.

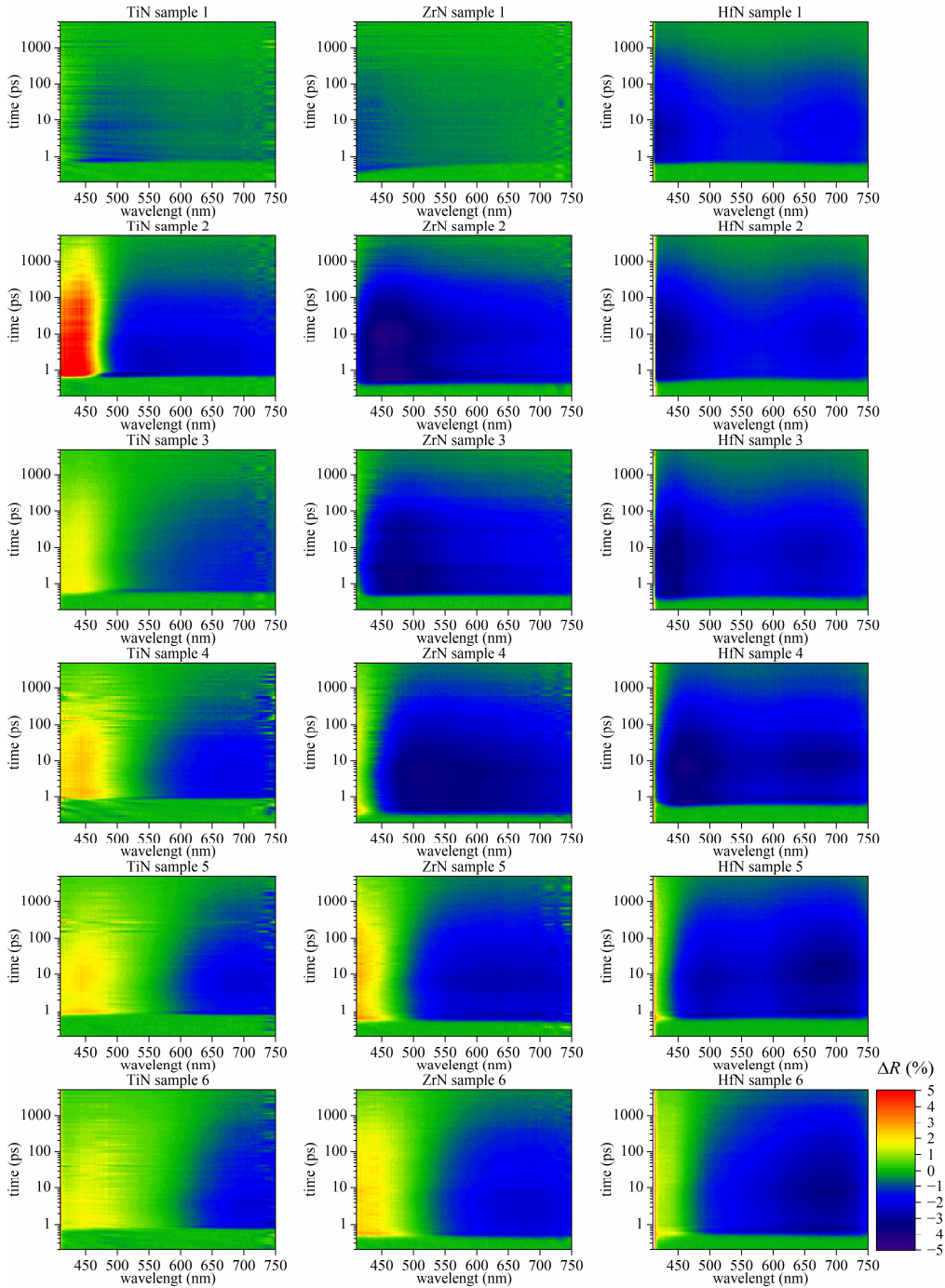

Fig. 2. Broadband transient reflectance maps of titanium, zirconium, and hafnium nitrides as a function of wavelength.

Figure 2 shows the broadband transient reflectance maps of titanium, zirconium, and hafnium nitrides as the function of wavelength (410 nm – 750 nm) and acquisition time (up to 5 ns). Some preliminary conclusions can be drawn by just qualitatively analyzing these maps. First, the observed transient changes are spectrally very broad. For example, the third zirconium nitride sample is characterized by negative changes in the reflectance smoothly and continuously covering the whole spectral measurement range, i.e., more than 350 nm, meaning, ZrN becomes less reflective. Second, for part of the results, especially those related to samples with higher nitrogen content and characterized by ENZ region in the measurement range of the ultrafast experimental setup, an area of opposite (here, positive) sign can be identified. These two areas (of positive and negative changes in reflectance) are separated by a narrow spectral region typically characterized by minor changes in transient reflectance upon ultrafast excitation. Third, the changes in the reflectance in the time domain are long-lasting. The reflectance does not fully recover for most samples, even after 5 ns, which is opposite to other ENZ materials like noble metals and transparent conductive oxides characterized by a recovery time in the picosecond time scale or less. We also note that an additional feature at approximately 700 nm can be identified for hafnium nitride samples. This feature can be attributed to an optical transition [34]. We also note that our internal test on titanium, zirconium, and hafnium nitride samples with the most nitrogen excess showed that the pump fluence does not influence ultrafast results qualitatively, both in the spectral and time domain.

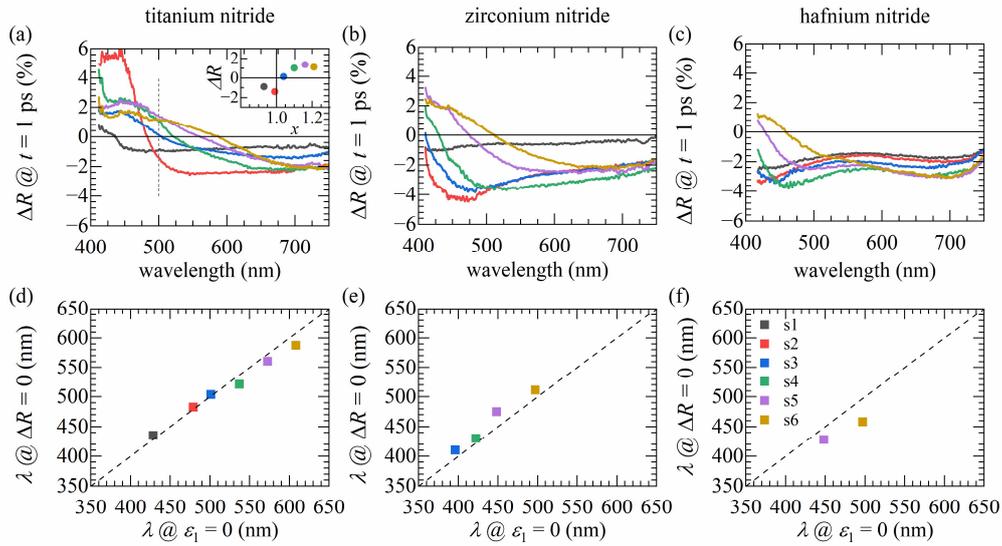

Fig. 3. The change in *R* at *t* = 1 ps as a function of stoichiometry for (a) TiN$_x$, (b) ZrN$_x$, and (c) HfN$_x$. Correlation between the wavelength at which Δ*R* equals zero and the wavelength at which the real $\varepsilon_1$ part of the dielectric function equals zero for (d) TiN$_x$, (e) ZrN$_x$, and (f) HfN$_x$. The inset in figure (a) shows changes in the reflectance taken at the time of 1 ps and the wavelength of 500 nm as a function of sample composition *x*.

To investigate the transient response of the refractory metal nitrides in more detail, we separately analyze the spectral and temporal dynamics. In this regard, Figures 3a–3c illustrate the spectral characteristics of the changes in transient reflectance Δ*R* at a selected time delay (*t* = 1 ps) for titanium, zirconium, and hafnium nitrides, respectively. Towards the blue wavelength of the ENZ region, the transient reflectance spectra report positive values at the excitation of pump pulses, as shown in Figure 3a–c. While reflectance decreases towards the red side of the ENZ region (metallic region). This means that the transient reflectance approaches zero value at the ENZ point and changes the sign from positive to negative values. All the nitride samples exhibit this response, but this behavior is partially visible in Zr and Hf nitride samples due to the absence of ultraviolet region A (315 nm – 400 nm) in the broadband probe beam of our ultrafast pump-probe set-up. Collation of particular wavelengths where transient reflectance equals zero λ@Δ*R*=0 and the wavelength at which the real part of the permittivity equals zero λ@$\varepsilon_1$=0 unveil a direct correlation, as shown in Figures 3d–3f. There are two critical outcomes based on these data.

First, it is possible to tune the spectral properties of the ultrafast response of the refractory metal nitrides by changing the samples' stoichiometry, which provides new functionality to this exciting class of optical materials. For example, in the case of titanium nitride for a selected wavelength of λ = 500 nm, the reflectance *R* upon femtosecond excitation after *t* = 1 ps can decrease for stoichiometric films, as well as for films with nitrogen deficiency (samples 1–3), can increase for films with higher nitrogen content (samples 5–7) and can stay constant for films with small nitrogen excess (sample 4). Thus, one can imagine, e.g., a metasurface comprising titanium nitride (among others) characterized by different compositions at different places on the surface, resulting in place-dependent changes (e.g., null, negative, or positive) in reflectance upon strong light illumination. Such a feature could increase freedom in shaping the optical properties of future advanced photonic devices comprising refractory metal nitrides.

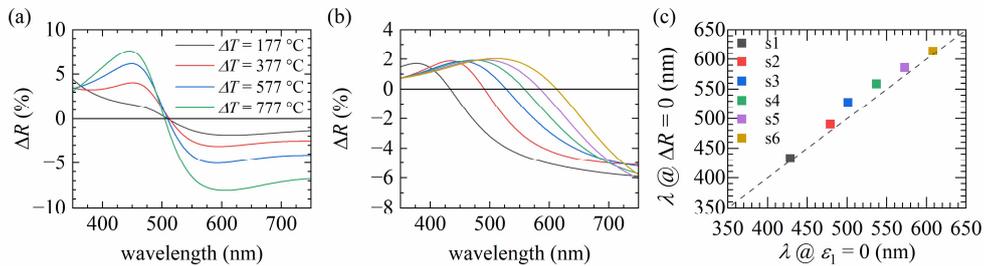

Figure 4. Calculated stationary changes in reflectance including only (a) the temperature-induced changes in the permittivity for stoichiometric samples (based on data taken from the literature [35]), (b) the increase in the Drude damping coefficient reflecting in a very simplified and limited manner increase in the temperature for stoichiometric and non-stoichiometric samples. (c) Correlation between the wavelength at which Δ*R* in (b) equals zero and the wavelength at which the real $\varepsilon_1$ part of the dielectric function equals zero.

Secondly, there is a correlation between the wavelength at which the changes in reflectance upon excitation equal zero λ@Δ*R*=0 and the wavelength at which the real part of the permittivity equals zero λ@$\varepsilon_1$=0. Since the changes in transient reflectance seem dominated by the thermal effect, we analyze how the thermally induced changes in permittivity would manifest themselves in the reflectance spectrum. For this purpose, we have taken literature data concerning the temperature-dependent permittivity of the 200 nm thick stoichiometric titanium nitride film [35] and calculated the expected changes in reflectance ($\Delta R$) (Figure 4a). The calculated static changes in reflectance qualitatively agree with our transient results; however, a small red-shift of the wavelength at which changes in reflectance approach towards zero with respect to the ENZ wavelength is observed. To understand the relation between the crossing point of the transient reflectance ($\lambda@\Delta R=0$) and the stoichiometry of the material, we performed a simple numerical experiment. First, we analytically parametrized the real and imaginary parts of the titanium nitride samples' permittivity using four Lorentz oscillators and one Drude component. Next, we calculated the reflectance spectrum $R$ for "unperturbed" titanium nitride semi-infinite films and for similar films but characterized by the Drude damping coefficient $\Gamma_D$ that was artificially increased by 50 %, which, in a very simplified and limited manner, reflects the effect of the increased temperature [36]. The results are shown in Figure 4b, whereas in Figure 4c, one can see the collation of the wavelength at which the changes in simulated reflectance equal zero and the wavelength at which the real part of the permittivity equals zero. As can be seen, surprisingly, it turned out that even so simple inclusion of temperature is enough to reproduce the observed correlation. Thus, it seems that temperature-induced changes in the permittivity result in changes in reflectance characterized by the crossing point correlated with the ENZ wavelength. If the ENZ wavelength changes, the crossing point will follow it. We want to underline at this point that the ENZ wavelength, characterizing the refractory metal nitrides, is not related directly to any optical transition but is a result of the interplay between the interband and intraband contributions to the permittivity that are well described by a set of Lorentz oscillators and the Drude component, respectively.

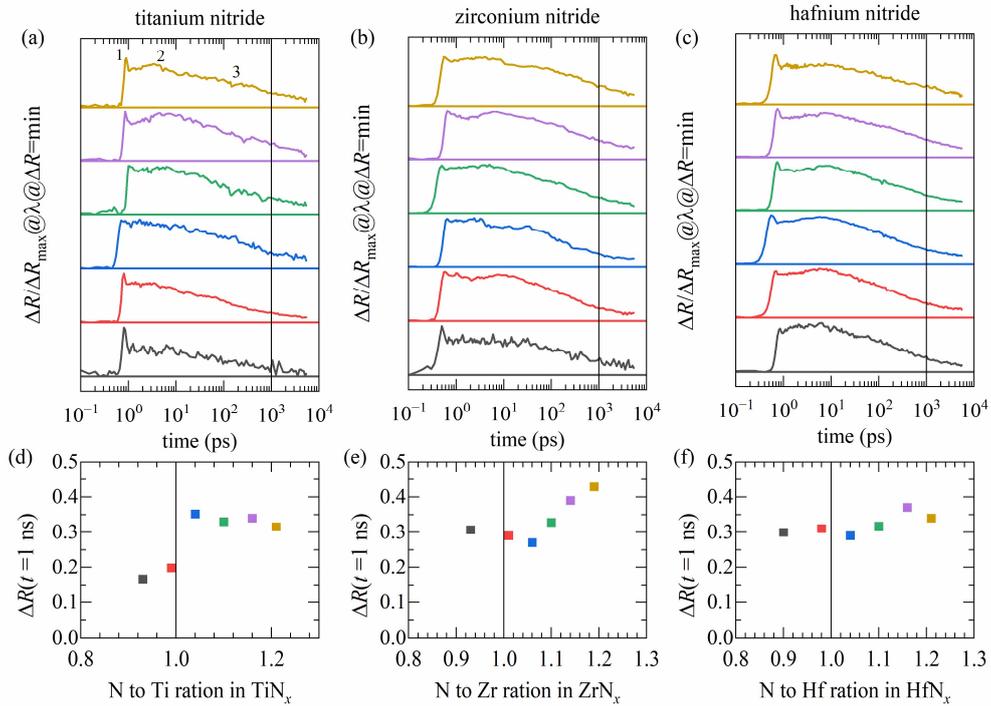

Figure 5. Normalized changes in $\Delta R$ at the wavelength at which the changes in $\Delta R$ are minimum for (a) TiN$_x$, numbers "1", "2", and "3" denote three components in the time response, (b) ZrN$_x$, and (c) HfN$_x$. Correlation between the kinetics of the changes in reflectance (normalized changes in $\Delta R$ at the wavelength at which the changes in $\Delta R$ are minimum taken at 1 ns) and nitrogen to the metal ratio for (d) TiN$_x$, (e) ZrN$_x$, and (f) HfN$_x$.

To shed some light on the changes in the reflectance in the time domain, we analyze the time dependence of the normalized changes in reflectance upon ultrafast illumination taken at one selected wavelength for each sample, as shown in Figures 5a–c. The selected wavelength is the one for which the changes in reflectance are the most negative. We do not take the data for a constant wavelength since the reflectance shifts in the spectral domain as the material and stoichiometry change.

In all the traces shown in Figures 5a–c, three components that comprise the time response can be identified. The first component is the tall, short peak (denoted "1" in Figure 5a), which renders the photoexcitation process related to the real intra and interband absorption as well as the following energy relaxation through the electron-electron and electron-phonon scattering mechanisms [37]. The decay time is deep in the subpicosecond scale, which agrees with the strong electron-phonon coupling hypothesis [38]. We note that we called this component "fast" to express the timescale (femtosecond), not the physical mechanism related to virtual electronic transitions, as often understood in nonlinear optics [39]. The second component (denoted "2") is related to the occurrence of the second maximum in the time response of the reflectance in the pico-second time scale. This second maximum has been recently identified as a manifestation of the thermoelastic effect [18]. Still, we note that although it was not named, it can also be observed in data reported in earlier works [11]. Interestingly, for the second and third hafnium nitride samples, the second maximum in the reflectance resulting from the thermoelastic effect is more pronounced than the first maximum resulting from the photogenerated hot carriers. Therefore, the thermoelastic effect can significantly affect the time response of the refractory metal nitrides, and it is not always only a small correction. The third and last component (denoted "3") is the long-lasting tail, which we attribute to the effect of the temperature-induced changes in the permittivity, which slowly diminishes in time due to the heat dissipation process. A significant thermal component in the time response of the refractory metal nitrides, which we report in this paper, is also in line with the opinion on these materials as very efficient light-heat transducers, as described in the introduction. Noticeable light absorption due to real electronic transitions, efficient energy transfer from the electronic to lattice subsystems, and significant local temperature increase are exactly these phenomena that can also explain our observations.

The last question we address in this paper is the dependence of the reflectance recovery time on stoichiometry. In Figures 5d–f, we plotted the normalized changes in reflectance from Figures 5a–c taken at 1 ns as the function of the nitrogen-to-metal ratio. First, we note that even after 1 ns, the changes in reflectance are still 17 % – 43 % of its maximum value. Second, the changes in reflectance do not drastically depend on stoichiometry; only some increases can be seen as the nitrogen-to-metal ratio increases, and only for titanium and zirconium nitrides. But what is important is the context of these observations. Our polycrystalline samples are deposited on a non-matched substrate at room temperature. Consequently, we expected that a significant number of defects would characterize all of them. Moreover, the deposition of non-stoichiometric films leads to distortion of the ideal crystal structure, e.g., by introducing interstitial atoms or vacancies, further increasing the already expected high amount of structural defects. So if the long tail were due to the hot carriers, we would expect a short recovery time (at least shorter than in the case of epitaxial films) and, in the case of non-stoichiometric samples, a further decrease in its value, analogously as in the case of, e.g., III-V semiconductors, where defects decrease the electron-hole recombination time due to the band gap's introduced energy levels, that form efficient recombination paths. On the other hand, if the long tail in the time response of the reflectance results from the heat dissipation process, we would expect that the influence of the nonstoichiometry on the recovery time would be minor due to the expected minor influence of the nonstoichiometry on the thermal conductivity of our polycrystalline films.

4. Conclusions

The results show that the thermal effect in the spectral and time domain can dominate the reflectance of refractory metal nitrides upon intense femtosecond light illumination. We demonstrate that by stoichiometry control, which from the technological point of view is a facile and reliable task, we control the spectral response of these materials. We show that changing stoichiometry makes it possible to achieve positive, negative, or minor changes in the reflectance at selected wavelengths upon intense femtosecond light illumination. This feature could increase freedom in shaping the optical properties of future advanced photonic devices comprising refractory metal nitrides. We also show that the stoichiometry changes do not significantly affect the long-lasting changes in reflectance, which not only supports the thesis on the thermal origin of the long tail but also allows for shaping the spectral response of the nitrides while keeping the time response on a similar level.

**Funding.** This work was supported by the POB FOTECH-3 project entitled "Plasmons and polaritons on nanostructured surfaces of IVb metal nitrides" granted by the Warsaw University of Technology within "The Excellence Initiative – Research University" program. We acknowledge the financial support of the European Research Council (Starting Grant project aQUARiUM; Agreement No. 802986), Academy of Finland Flag-ship Programme (PREIN), (320165).

**Disclosures.** The authors declare no conflicts of interest.

**Data availability.** Data underlying the results presented in this paper are not publicly available at this time but may be obtained from the authors upon reasonable request.

**References**

1. G. V. Naik, J. Kim, and A. Boltasseva, "Oxides and nitrides as alternative plasmonic materials in the optical range [Invited]," Opt. Mater. Express, OME 1(6), 1090–1099 (2011).

2. G. V. Naik, V. M. Shalaev, and A. Boltasseva, "Alternative Plasmonic Materials: Beyond Gold and Silver," Advanced Materials 25(24), 3264–3294 (2013).

3. U. Guler, J. C. Ndukaife, G. V. Naik, A. G. A. Nnanna, A. V. Kildishev, V. M. Shalaev, and A. Boltasseva, "Local Heating with Lithographically Fabricated Plasmonic Titanium Nitride Nanoparticles," Nano Lett. 13(12), 6078–6083 (2013).

4. S. Ishii, R. P. Sugavaneshwar, and T. Nagao, "Titanium Nitride Nanoparticles as Plasmonic Solar Heat Transducers," J. Phys. Chem. C 120(4), 2343–2348 (2016).

5. L. Gui, S. Bagheri, N. Strohfeldt, M. Hentschel, C. M. Zgrabik, B. Metzger, H. Linnenbank, E. L. Hu, and H. Giessen, "Nonlinear Refractory Plasmonics with Titanium Nitride Nanoantennas," Nano Lett. 16(9), 5708–5713 (2016).

6. S. Ishii, S. L. Shinde, W. Jevasuwan, N. Fukata, and T. Nagao, "Hot Electron Excitation from Titanium Nitride Using Visible Light," ACS Photonics 3(9), 1552–1557 (2016).

7. A. Naldoni, U. Guler, Z. Wang, M. Marelli, F. Malara, X. Meng, L. V. Besteiro, A. O. Govorov, A. V. Kildishev, A. Boltas-seva, and V. M. Shalaev, "Broadband Hot-Electron Collection for Solar Water Splitting with Plasmonic Titanium Ni-tride," Advanced Optical Materials 5(15), 1601031 (2017).

8. R. A. Karaballi, Y. Esfahani Monfared, and M. Dasog, "Photothermal Transduction Efficiencies of Plasmonic Group 4 Metal Nitride Nanocrystals," Langmuir 36(18), 5058–5064 (2020).

9. P. Das, B. Biswas, K. C. Maurya, M. Garbrecht, and B. Saha, "Refractory Plasmonic Hafnium Nitride and Zirconium Nitride Thin Films as Alternatives to Silver for Solar Mirror Applications," ACS Appl. Mater. Interfaces 14(41), 46708–46715 (2022).

10. A. Pianelli, R. Dhama, J. Judek, R. Mazur, and H. Caglayan, "Two-color all-optical switching in Si-compatible epsilon-near-zero hyperbolic metamaterials," https://doi.org/10.48550/arXiv.2305.06731 (2023).

11. S. Saha, B. Diroll, M. G. Ozlu, S. N. Chowdhury, S. Peana, Z. Kudyshev, R. Schaller, Z. Jacob, V. M. Shalaev, A. V. Kildishev, and A. Boltasseva, "Engineering the Temporal Dynamics with Fast and Slow Materials for All-Optical Switch-ing," https://doi.org/10.48550/arXiv.2208.12927 (2022).

12. C.-K. Sun, F. Vallée, L. H. Acioli, E. P. Ippen, and J. G. Fujimoto, "Femtosecond-tunable measurement of electron thermalization in gold," Phys. Rev. B 50(20), 15337–15348 (1994).

13. J. Hodak, I. Martini, and G. V. Hartland, "Ultrafast study of electron–phonon coupling in colloidal gold particles," Chem-ical Physics Letters 284(1–2), 135–141 (1998).

14. S. Link, C. Burda, Z. L. Wang, and M. A. El-Sayed, "Electron dynamics in gold and gold–silver alloy nanoparticles: The influence of a nonequilibrium electron distribution and the size dependence of the electron–phonon relaxation," The Journal of Chemical Physics 111(3), 1255–1264 (1999).

15. N. Kinsey, C. DeVault, J. Kim, M. Ferrera, V. M. Shalaev, and A. Boltasseva, "Epsilon-near-zero Al-doped ZnO for ultra-fast switching at telecom wavelengths," Optica 2(7), 616 (2015).

16. M. Z. Alam, I. De Leon, and R. W. Boyd, "Large optical nonlinearity of indium tin oxide in its epsilon-near-zero region," Science 352(6287), 795–797 (2016).

17. H. George, J. Reed, M. Ferdinandus, C. DeVault, A. Lagutchev, A. Urbas, T. B. Norris, Vladimir. M. Shalaev, A. Boltasseva, and N. Kinsey, "Nonlinearities and carrier dynamics in refractory plasmonic TiN thin films," Opt. Mater. Express 9(10), 3911 (2019).

18. B. T. Diroll, S. Saha, V. M. Shalaev, A. Boltasseva, and R. D. Schaller, "Broadband Ultrafast Dynamics of Refractory Metals: TiN and ZrN," Adv. Optical Mater. 8(19), 2000652 (2020).

19. S. Rotta Loria, B. R. Bricchi, A. Schirato, L. Mascaretti, C. Mancarella, A. Naldoni, A. Li Bassi, G. Della Valle, and M. Zavelani-Rossi, "Unfolding the Origin of the Ultrafast Optical Response of Titanium Nitride," Advanced Optical Materials 2300333 (2023).

20. S. Chung, S. Shrestha, X. Wen, Y. Feng, N. Gupta, H. Xia, P. Yu, J. Tang, and G. Conibeer, "Hafnium nitride for hot carrier solar cells," Solar Energy Materials and Solar Cells 144, 781–786 (2016).

21. H. Xia, X. Wen, Y. Feng, R. Patterson, S. Chung, N. Gupta, S. Shrestha, and G. Conibeer, "Hot carrier dynamics in HfN and ZrN measured by transient absorption spectroscopy," Solar Energy Materials and Solar Cells 150, 51–56 (2016).

22. S. Chung, X. Wen, S. Huang, N. Gupta, G. Conibeer, S. Shrestha, T. Harada, and T. W. Kee, "Nanosecond long excited state lifetimes observed in hafnium nitride," Solar Energy Materials and Solar Cells 169, 13–18 (2017).

23. D. B. O'Neill, S. K. Frehan, K. Zhu, E. Zoethout, G. Mul, E. C. Garnett, A. Huijser, and S. H. C. Askes, "Ultrafast Photoinduced Heat Generation by Plasmonic HfN Nanoparticles," Advanced Optical Materials 9(19), 2100510 (2021).

24. A. Bendavid, P. J. Martin, R. P. Netterfield, and T. J. Kinder, "Characterization of the Optical Properties and Composition of TiNx Thin Films by Spectroscopic Ellipsometry and X-ray Photoelectron Spectroscopy," Surface and Interface Analy-sis 24(9), 627–633 (1996).

25. C. G. H. Walker, J. A. D. Matthew, C. A. Anderson, and N. M. D. Brown, "An estimate of the electron effective mass in titanium nitride using UPS and EELS," Surface Science 412–413, 405–414 (1998).

26. J. H. Kang and K. J. Kim, "Structural, optical, and electronic properties of cubic TiNx compounds," Journal of Applied Physics 86(1), 346–350 (1999).

27. J. Judek, P. Wróbel, P. P. Michałowski, M. Ożga, B. Witkowski, A. Seweryn, M. Struzik, C. Jastrzębski, and K. Zberecki, "Titanium Nitride as a Plasmonic Material from Near-Ultraviolet to Very-Long-Wavelength Infrared Range," Materials 14(22), 7095 (2021).
28. H. M. Benia, M. Guemmaz, G. Schmerber, A. Mosser, and J. C. Parlebas, "Optical properties of non-stoichiometric sputtered zirconium nitride films," Applied Surface Science 211(1–4), 146–155 (2003).
29. R. Lamni, E. Martinez, S. G. Springer, R. Sanjinés, P. E. Schmid, and F. Lévy, "Optical and electronic properties of magnetron sputtered ZrNx thin films," Thin Solid Films 447–448, 316–321 (2004).
30. Q. Guo, T. Wang, Y. Ren, Y. Ran, C. Gao, H. Lu, Z. Jiang, and Z. Wang, "Plasmonic properties of nonstoichiometric zirconium nitride, oxynitride thin films, and their bilayer structures," Phys. Rev. Materials 5(6), 065201 (2021).
31. J. Bulíř, J. More-Chevalier, S. Chertopalov, L. Fekete, L. Volfová, P. Hubík, M. Novotný, and J. Lančok, "Analysis of thickness-dependent electron transport in magnetron sputtered ZrN films by spectroscopic ellipsometry," Thin Solid Films 731, 138746 (2021).
32. C. Hu, Z. Gu, J. Wang, K. Zhang, X. Zhang, M. Li, S. Zhang, X. Fan, and W. Zheng, "Nature of Tunable Optical Reflectivity of Rocksalt Hafnium Nitride Films," J. Phys. Chem. C 118(35), 20511–20520 (2014).
33. S. H. C. Askes, N. J. Schilder, E. Zoethout, A. Polman, and E. C. Garnett, "Tunable plasmonic HfN nanoparticles and arrays," Nanoscale 11(42), 20252–20260 (2019).
34. B. Saha, J. Acharya, T. D. Sands, and U. V. Waghmare, "Electronic structure, phonons, and thermal properties of ScN, ZrN, and HfN: A first-principles study," Journal of Applied Physics 107(3), 033715 (2010).
35. H. Reddy, U. Guler, Z. Kudyshev, A. V. Kildishev, V. M. Shalaev, and A. Boltasseva, "Temperature-Dependent Optical Properties of Plasmonic Titanium Nitride Thin Films," ACS Photonics 4(6), 1413–1420 (2017).
36. J. A. Briggs, G. V. Naik, Y. Zhao, T. A. Petach, K. Sahasrabuddhe, D. Goldhaber-Gordon, N. A. Melosh, and J. A. Dionne, "Temperature-dependent optical properties of titanium nitride," Appl. Phys. Lett. 110(10), 101901 (2017).
37. A. Schirato, M. Maiuri, G. Cerullo, and G. Della Valle, "Ultrafast hot electron dynamics in plasmonic nanostructures: experiments, modelling, design," Nanophotonics 12(1), 1–28 (2023).
38. S. Dal Forno and J. Lischner, "Electron-phonon coupling and hot electron thermalization in titanium nitride," Phys. Rev. Materials 3(11), 115203 (2019).
39. J. B. Khurgin, M. Clerici, and N. Kinsey, "Fast and Slow Nonlinearities in Epsilon-Near-Zero Materials," Laser Photonics Rev. 15, 2000291 (2021).